\input{psfig}
\documentstyle[11pt]{article}
\newcommand{\be}{\begin{equation}}
\newcommand{\ee}{\end{equation}}
\newcommand{\ba}{\begin{eqnarray}}
\newcommand{\ea}{\end{eqnarray}}

\begin{document}

\title{\vspace{-3.0cm} \hspace{0.0cm} \hspace*{\fill} \\[-5.5ex]
\hspace*{\fill}{\normalsize LA-UR-96-2868} \\[1.5ex]
{\huge {\bf On the m$_\perp$-dependence of Bose-Einstein
correlation radii}}}

\author{B.R. Schlei${}^1$\thanks{E. Mail: schlei@t2.LANL.gov}{\ } and
N. Xu${}^2$\thanks{E. Mail: nu\_xu@lanl.gov}\\[1.5ex]
{\it ${}^1$Theoretical Division, Los Alamos National Laboratory,
Los Alamos, NM 87545, USA}\\
{\it ${}^2$P-25, Los Alamos National Laboratory,
Los Alamos, NM 87545, USA}
}
\date{\today}
\maketitle

\begin{abstract} Within a theoretical study using both HYLANDER 
and RQMD, we revisit Bose-Einstein correlation measurements of 
200AGeV S+S and 160AGeV Pb+Pb. Transverse flow does not show up in the 
Bose-Einstein measurements. The decrease of the effective transverse radius
parameters with increasing transverse average momentum of the particle
pair is largely due to the effect of resonance decays.\end{abstract}

\hbadness=10000
\vspace{-0.5cm}
\newpage

In the ongoing search for the quark-gluon plasma (QGP), Bose-Einstein
correlations (BEC) of identical bosons have become an issue of great
current interest. BEC serve as a tool in the determination of radii
(spatial extensions) and lifetimes (temporal extensions) of hadron
emission zones generated in relativistic nucleus-nucleus collisions
\cite{boal}.  However, the inverse widths of BEC functions do not have
a simple interpretation in terms of those radii and lifetimes. The inverse 
widths or ``correlation radii'' have to be considered as combinations of 
space-time moments from space-time densities ({\it cf.}, e.g., ref. 
\cite{bernd9} and references therein). The phase-space distribution, or 
the source function $g(x,p)$, does not only depend on the space-time 
point $x$ and the four-momentum $p$ of the emitted particles, but also 
on the temporal evolution of the hadron source (fireball) and, more 
importantly, on the correlations among $x$ and $p$ \cite{pratt}. Furthermore, 
it is known that BEC are sensitive to the effects of the decay of unstable 
particles (resonances). Experimentally obtained BEC functions are always 
averaged over certain phase-space regions and are therefore in general 
sensitive to the specific kinematical regions under consideration. 
Because of the complications involved in BEC, it is necessary to account 
for all known effects before we can draw conclusions about new physics which
might emerge in relativistic nucleus-nucleus collisions.

Recent theoretical studies \cite{csorgo2}-\cite{cs9509} claim that the
inverse widths of BEC functions in relativistic heavy-ion collisions
at CERN/SPS energies near 200 GeV per nucleon show a $m_\perp$-scaling
behavior, namely that the size parameters are inversely proportional to
the square-root of the transverse mass, $\sqrt{m_{\perp}}$. This idea was 
proposed earlier by Sinyukov et al. \cite{sinyukov} in 1987, in which the
authors suggested that the longitudinal size parameter, $R_\parallel$, 
should follow the $m_\perp$-scaling:
\begin{equation}
R_\parallel \propto \sqrt{\frac{T_f}{m_\perp}}\frac{1}{\cosh(y_K)} \:,
\label{eq:rlong}
\end{equation}

where $T_f$ and $y_K$ are the particle freeze-out temperature and rapidity of 
the particle pair, respectively.  The pair transverse mass is defined as 
$m_\perp=\sqrt{m^2+\vec{K}_\perp^2}$ with the average transverse momentum 
$\vec{K}_\perp=(\vec{p}_{1\perp}+\vec{p}_{2\perp})/2$ of the particle pair. 
This relationship has been indeed observed in the longitudinal size parameter 
$R_\parallel$ from the sulphur-induced high energy collisions \cite{QM95} and 
the effect is attributed to the collective expansion. Neglecting the resonance
decays, the authors in refs. \cite{csorgo2}-\cite{cs9509} suggested that even 
in the transverse directions, size parameters $R_{out}$ and $R_{side}$ 
also follow the same $m_{\perp}$-scaling law. On the other hand, it has been 
shown that the effect of the resonance decay strongly affect the inverse 
widths of the BEC \cite{bernd2}-\cite{bernd7}.

It is the purpose of this short note to study the $m_\perp$-scaling
behaviour by using two realistic source models: HYLANDER \cite{udo} and
RQMD \cite{rqmd1}. This study includes effects of transverse expansion and 
decay of resonances on BEC of identical particles. We shall first compare
HYLANDER results with recent measurements of S+S and Pb+Pb collisions. 
Then we will focus on the $m_\perp$-scaling issues by inspecting the 
$m_\perp$-dependence of the size parameters calculated from both models.

The HYLANDER model belongs to the class of models applying
(3+1)-dimensional relativistic one-fluid-dynamics (for other
hydrodynamical models {\it cf.} \cite{bernd8} and refs. therein).
HYLANDER provides fully three-dimensional numerical solutions of the
hydrodynamical relativistic Euler-equations \cite{euler}. The model
has been successfully applied to several different heavy-ion
collisions at SPS energies \cite{marburg}. Once the initial conditions
and the equation of state are specified, one obtains an unambiguous
solution from the hydrodynamical equations.  The choice of a
freeze-out condition such as, {\it e.g.}, that the expanding fluid
reaches a fixed freeze-out temperature, $T_f$, determines the final
space-time geometry of the hydrodynamically expanding fireball.  In
particular, for 200AGeV S+S and 160AGeV Pb+Pb central collisions, the
calculation of single inclusive spectra of negative hadrons and
protons and BEC of negative pions and kaons was performed using the
formalism outlined in refs.  \cite{bernd2,bernd3,bernd1}. 
Resonance decays were included in these calculations. The results for
BEC of negative pions and kaons have been published in refs.
\cite{bernd2}-\cite{bernd7},\cite{bernd8}.

In order to extract effective hadron source radii, the results of the
BEC calculations have been fitted to the Gaussian form\footnote{If one 
performs a fit to a BEC function in more than one dimension, eq. 
(\ref{eq:fit}) does not represent the most general expression, because of the 
existence of an ``out-longitudinal'' cross term \cite{chapman}.}
which has been widely used by experimentalists for the
presentation of their BEC data (for the choice of the variables, {\it
cf.} ref. \cite{bertsch}):
\begin{equation}
C_2(\vec{p}_1,\vec{p}_2)\:=\: 1\:+\:\lambda\:\exp
\left[-\frac{ 1}{2} \left(q_\parallel^2 R_\parallel^2(\vec{K}) \:+\:q_{side}^2
R_{side}^2(\vec{K})\:+\:q_{out}^2 R_{out}^2(\vec{K})\right) \right] \:.
\label{eq:fit}
\end{equation} 

In eq. (\ref{eq:fit}) $\vec{p}_1$, $\vec{p}_2$ are the momenta of the two 
emitted identical particles, whereas $\vec{K}$ denotes the average momentum 
of the particle pair. The $q_i$ ($i={\parallel}, {side}, {out}$) are
components of the momentum difference vector $\vec{q}$ of the particle pair.
It should be emphasized that in the present model $\lambda \equiv
\lambda (\vec{K})$ does {\it not} represent the effect of coherence, but the 
momentum-dependent effective reduction of the intercept due to the 
contributions from the decay of long-lived resonances (such as 
$\eta$, $\eta^\prime$ ... etc.) \cite{bernd2}-\cite{bernd7},\cite{bernd8}.
A characteristic property of particle production from an expanding source
is a correlation between the space-time point where a particle is emitted
and its energy-momentum \cite{pratt}. As a consequence, the inverse widths
$R_i(\vec{K})$ ($i={\parallel}, {side}, {out}$) extracted from 
Bose-Einstein correlation functions show a characteristic dependence of the 
average momentum of the pair, $\vec{K}$, and are therefore dependent on
detector acceptances.

Fig. 1 (solid lines) shows the calculations for the effective radii
$R_\parallel$, $R_{side}$ and $R_{out}$ as functions of the transverse
momentum $K_\perp$ of the pion pair compared to the corresponding NA35
and preliminary NA49 data \cite{QM95,alber}, respectively. In order to
make a comparison of the calculated effective radii with the
experimentally obtained ones, detector acceptances have been accounted
for.  In the case of 200AGeV S+S the effective radii have been calculated
as functions of $K_\perp$ at the rapidity of the pair 
$y_K=4.0-y_{cm} \approx 1.0$ . In the case of 160AGeV Pb+Pb
the effective radii were calculated as functions of $K_\perp$ at
$y_K=4.5-y_{cm} \approx 1.6$.  The effective longitudinal radii
$R_\parallel$ are evaluated in the longitudinal comoving system (LCMS),
with $\gamma_\parallel = \cosh(y_K)$. All of these calculations, which
in the case of S+S had been true predictions, agree surprisingly
well with the data.

The effective radii of only the directly emitted (thermal) negative
pions, $\pi^-$, (dashed lines) are also plotted in Fig. 1.  We can see
that the effective radii of the directly emitted $\pi^-$
are smaller compared to those of all $\pi^-$, {\it i.e.},
including the pions originating from resonance decays, and follow a
{\it different} $K_\perp$-dependence. The effective radii for all
pions are larger compared to the effective radii of directly emitted
pions, because the unstable particles (resonances) have a finite
lifetime. Within their lifetime the resonances can propagate to a
distant location from their points of origin (the fireball). While
decaying the resonances generate a cloud of pion emission locations
around the actual fireball and therefore increase the effective size
of the system.

Fig. 1 shows that the $K_\perp$-dependence of the effective transverse
radii of thermal pions is very weak while the effective transverse radii 
of all pions significantly decrease as a function of $K_\perp$. In refs.
\cite{bernd7,bernd8} it was explained that the hydrodynamical solutions
express transverse expansion. The two systems each have a maximum
value for the transverse velocities: for 200AGeV S+S we have a maximum
transverse velocity $u_\perp^{max}(S)=0.43$ whereas for 160AGeV Pb+Pb we 
have $u_\perp^{max}(Pb)=0.61$. The maximal values for the transverse 
velocities correspond to fluid cells which have space-time rapidities 
$\eta \approx 0.5 \cdot \ln(t+z/t-z)$ equal to zero ($t$ and $z$ are the 
time and the longitudinal spacial coordinate, respectively). 
On the contrary, fluid cells with large space-time rapidity have very small 
corresponding values for the transverse velocities at freeze-out ({\it cf.} 
Fig. 6 in ref. \cite{bernd8}). We have checked that in the phase-space regions,
where the BEC data of the NA35 and NA49 Collaborations have been taken, the 
contributions to the BEC functions come mainly from fluid cells with almost 
vanishing transverse velocity at freeze-out.  

Transverse flow has two major effects on effective BEC radii. First there is 
an apparent reduction of the effective longitudinal radii, because the fluid 
also expands transversely. Secondly, the transverse effective radii show a
$K_\perp$-dependence. For the particular hydrodynamical solutions
under consideration transverse flow results in a decrease of the
transverse effective radii due to the effects of relatively  fast transverse
inwardly moving rarefaction waves. In agreement with ref. \cite{bernd1}, the 
effective transverse radii of directly emitted negative pions plotted in 
Fig.1 show almost no $K_\perp$-dependence. But the effective radii of all
$\pi^-$ have a $K_\perp$-dependence which would indicate a present
transverse flow in the data. Since there is no transverse flow in the
considered phase-space region as we have argued above, the decrease of
the effective radii for all $\pi^-$ must have a different reason. In
fact, for smaller transverse momenta we have relatively larger
contributions from resonance decay to the BEC than at higher transverse 
momenta $K_\perp$. Since the relative weight of pions being produced
directly compared to those originating from resonance decays enters
in the composition of BEC functions, the decrease of the effective
radii of all pions with increasing $K_\perp$ can be understood in
terms of the chemical composition at freeze-out as a function of
particle momenta. Resonance decays produce a pionic cloud that surrounds
the fireball. This pion cloud has in general larger spatial extensions 
(and a longer lifetime) than the fireball which emits the direct pions. 
Thus relatively larger resonance contributions result in relatively larger 
correlation radii and therefore, the decay of resonances is responsible for 
the decrease of the transverse radii with increasing $K_\perp$.

Let us now discuss $m_\perp$-scaling according to refs. 
\cite{csorgo2}-\cite{cs9509}. If $m_\perp$-scaling would be present, 
we would get
\begin{equation}
\gamma_\parallel^2 R_\parallel^2(\vec{K}) \:\simeq\:R_{side}^2(\vec{K})
\:\simeq\:
R_{out}^2(\vec{K})\:\propto\:\displaystyle{\frac{T_f}{m_\perp}}.
\label{eq:scal}
\end{equation} 

The HYLANDER BEC calculations for the transverse radii $R_{out}$ and 
$R_{side}$ show an even stronger decrease with $K_\perp$ at midrapidity,
i.e., $y_K=0.0$ ({\it cf.} Fig. 2 in ref. \cite{bernd2} and Fig. 3a in ref. 
\cite{bernd7}). In order to get a maximum effect due to a possible
$m_\perp$-scaling we show in Fig. 2 effective radii calculated at $y_K=0.0$
and multiplied with $\sqrt{m_\perp/T_f}$ as a function of the pair 
transverse mass $m_{\perp }$. A present $m_\perp$-scaling would result in 
a constant behaviour of the plotted quantities, {\it i.e.}, there would be 
no change with changing $m_\perp$. Obviously, the effective radii multiplied 
with the factor $\sqrt{m_\perp/T_f}$ do not show an independence of $m_\perp$.
Consequently, the effective radii $R_\parallel$, $R_{side}$ and $R_{out}$ 
(and those of only directly emitted $\pi^-$) are not consistent with a simple 
$m_\perp$-scaling behaviour in this hydrodynamical model.

Up to now, we have based all of our arguments on the hydrodynamical
model HYLANDER and simple $m_{\perp}$-scaling is found neither in the
longitudinal direction, $R_\parallel$, nor in the transverse directions,
$R_{out}$ and $R_{side}$. In addition, we found that resonance decays
increase the size parameters at low transverse momenta dramatically.  

To further understand the physics of the transverse momentum
dependence of the size parameters, we made an additional test with the
widely used transport model RQMD (v2.2) \cite{sorge}. Using the
freeze-out phase-space distribution, $g(x,p)$, we calculated
the averaged pion size parameters $R_\parallel$, $R_{out}$, and
$R_{side}$ as a function of the pair transverse mass $m_{\perp}$ for
160AGeV Pb+Pb central collisions, multiplied with $\sqrt{m_\perp/T_f}$
and again at midrapidity, $y_K=0.0$. 
The results are shown in Fig. 3 where the solid lines represent all pions 
and dotted lines are for rescattered pions. Rescattered pions do not
originate from resonance decay at the moment of their last interaction. 
A temperature parameter of $T_f = 140 MeV$ is used here \cite{sorge}.
The rescattered pions are similar to the thermal pions 
from the hydrodynamical model calculations (see dashed lines in Figs. 1,2),
because they are emitted earlier.

It is well known by now that, due to rescatterings in the heavy-ion
collisions, collective flow manifests itself in the cascade type of
calculations \cite{xu1,xu2}. In the Pb+Pb central collisions, the
averaged transverse flow velocity at midrapidity is about 40\% of the
speed of light.  If the $m_\perp$-scaling \cite{csorgo2} is correct,
one would expect a flat distribution of the scaled size parameters
$\sqrt{m_\perp/T_f } R_i$ ($ i=\parallel, out, side$) with respect the
pair transverse mass, $m_{\perp}$. However, Fig. 3 does not
show such a strong flow effect as proposed in ref. \cite{csorgo2}.
It is worth noting, that in a recent publication \cite{na44} on S+Pb central 
collisions one indeed observed a $m_\perp$-scaling. The corresponding RQMD 
predictions which accounted for the exact experimental acceptance were 
consistent with the data. However, the experimental acceptance has a rapidity 
span larger than one unit and, at each transverse momentum, the corresponding 
rapidity is different. 
Such an experimental effect may explain the different $m_\perp$-behaviour
between the experimental results of NA35 \cite{QM95} and NA44 \cite{na44} 
and also the difference between our results in Fig. 3 and NA44.
 
To summarize, we have revisited Bose-Einstein correlation measurements for
200AGeV S+S and 160AGeV Pb+Pb.  For the theoretical analysis we have
used a hydrodynamical model (HYLANDER) and a cascade model (RQMD). Within 
the experimental acceptance, the hydrodynamical model results are consistent 
with the measured size parameters for both 200 AGeV S+S and 160 AGeV Pb+Pb
central collisions. However, at midrapidity, no $m_{\perp}$-scaling is
seen in the results from both model calculations. Resonance decays seem to
play an important role in the Bose-Einstein correlation radii.
Our conclusion is also consistent with recent publications 
\cite{wiedemann,heinz}.\\

We would like to thank D. Strottman for many helpful discussions and
for reading the manu\-script carefully. N.X. thanks Professor Yuri I. Sinyukov 
and Dr. Heinz Sorge for many exciting discussions on this matter. 
This work has been supported by the Departement of Energy.

\newpage
\noindent

{\Large {\bf Figure Captions}}

\begin{description}

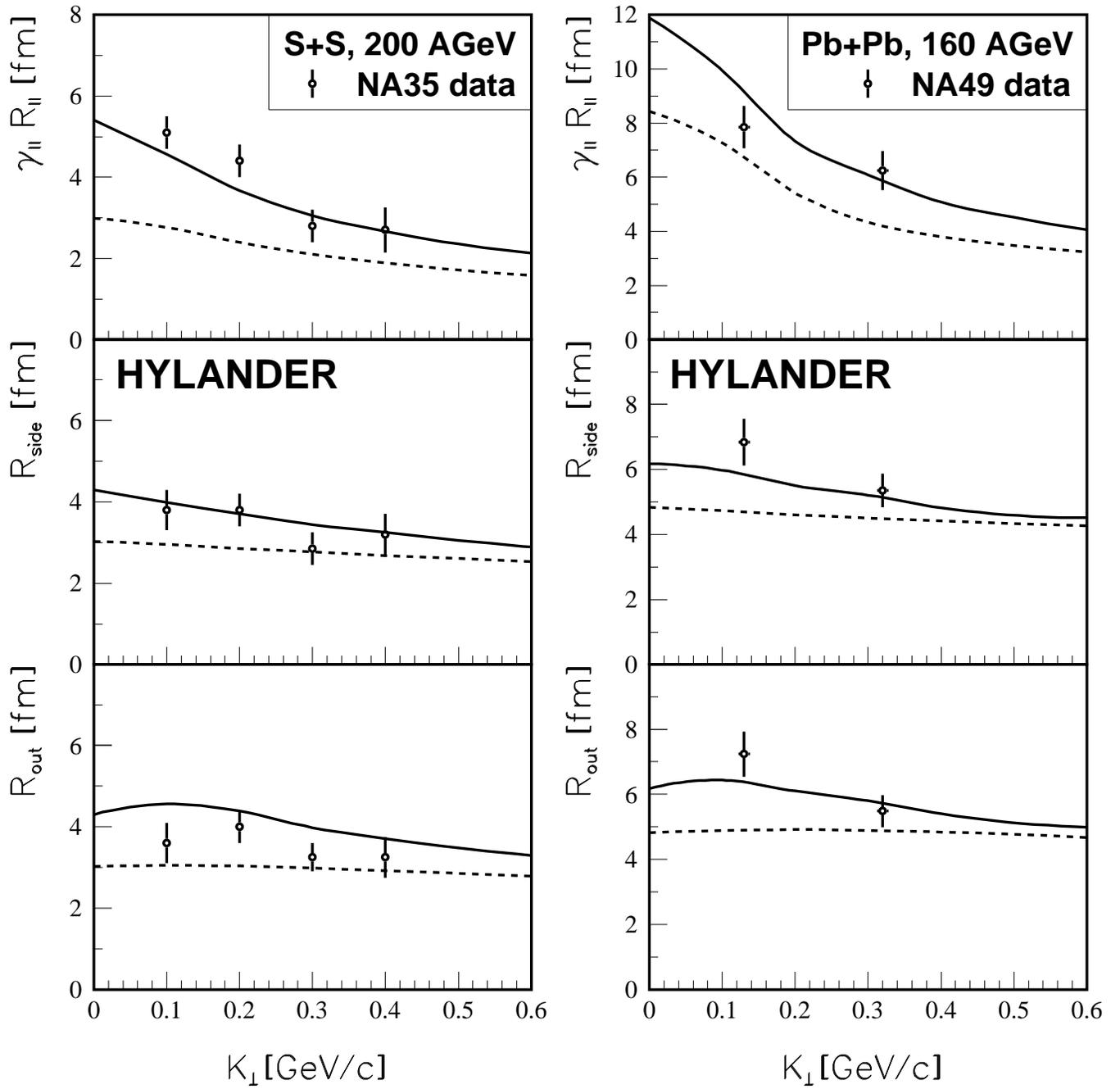
\item[Fig. 1] Effective radii extracted from Bose-Einstein correlation
functions of identical pions as a function of the transverse average
momentum $K_{\perp }$ of the pion pair at rapidity $y_K=4.0-y_{cm}
\approx 1.0$ for 200AGeV S+S and at rapidity $y_K=4.5-y_{cm}
\approx 1.6$ for 160AGeV Pb+Pb, respectively. Solid lines
correspond to all $\pi^-$, {\it i.e.}, including those from resonance
decay, whereas the dashed lines correspond to directly emitted
(thermal) $\pi^-$.  The theoretical results are compared to NA35 data
\cite{QM95,alber} and preliminary NA49 data \cite{QM95}.

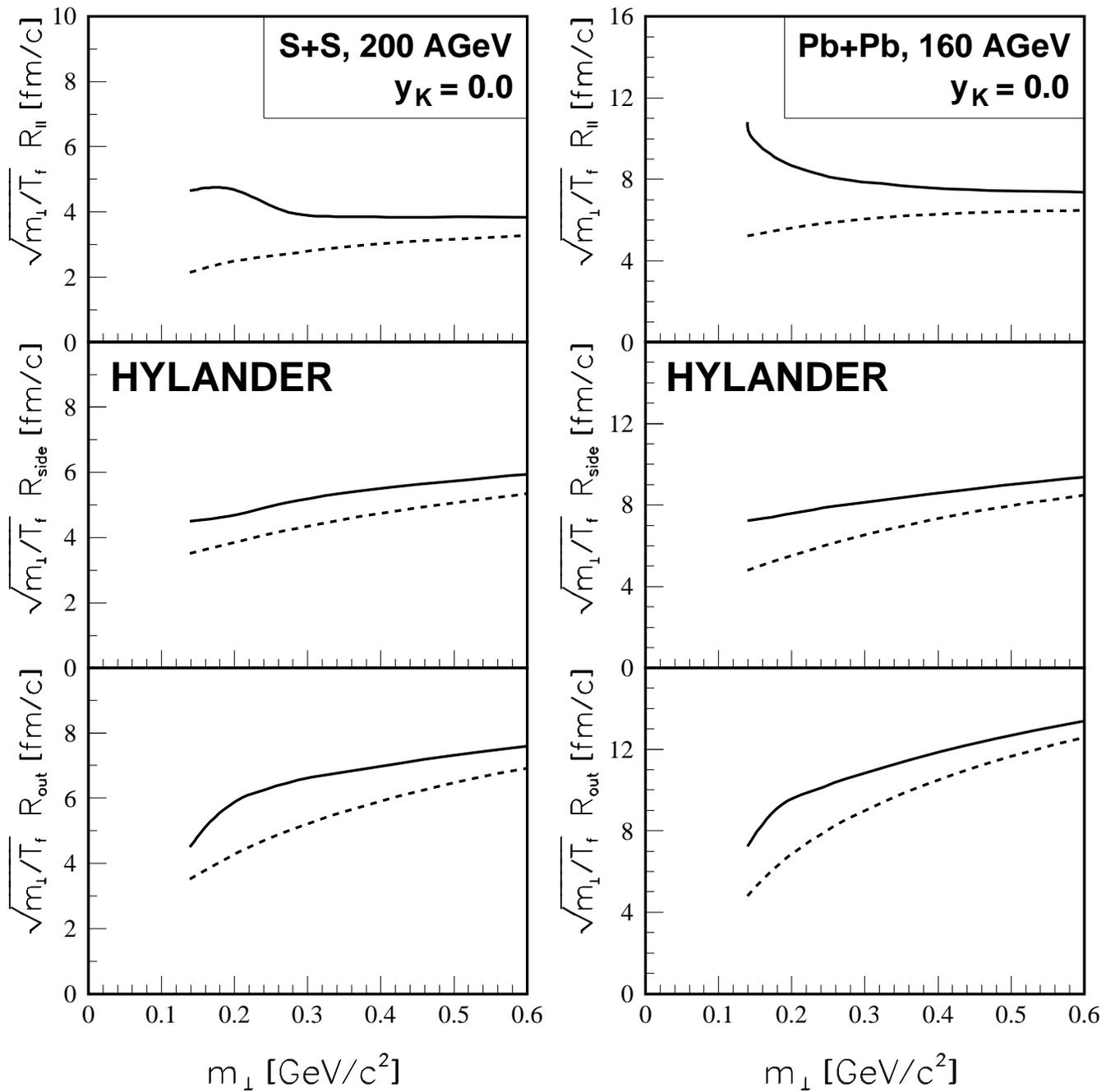
\item[Fig. 2] Effective radii extracted from Bose-Einstein correlation
functions of identical pions multiplied with $\sqrt{m_\perp/T_f}$ as a
function of the transverse mass $m_{\perp }$ of the pion pair at rapidity 
$y_K=0.0$ for 200AGeV S+S and for 160AGeV Pb+Pb, respectively. Solid lines 
correspond to all $\pi^-$, {\it i.e.}, including those from resonance decay, 
whereas the dashed lines correspond to directly emitted (thermal) $\pi^-$
only.  

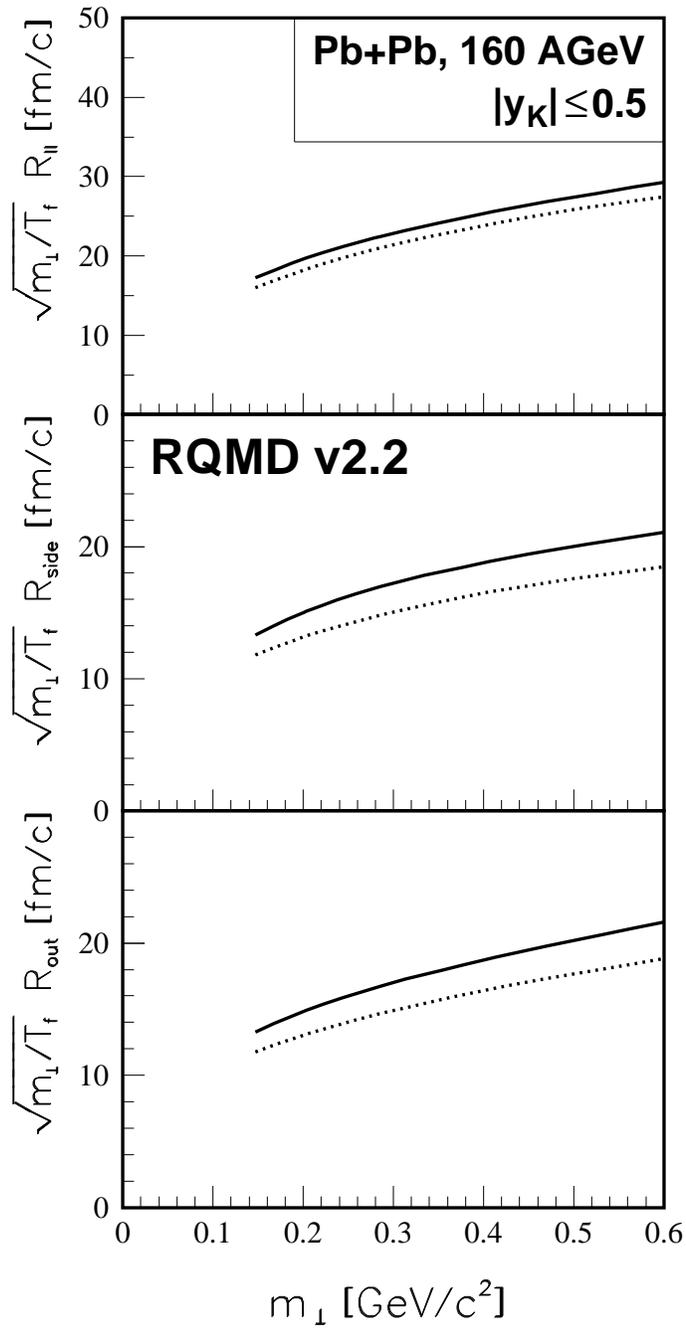
\item[Fig. 3] Scaled radii from the transport model RQMD (v2.2). 
A freeze-out temperature of 140 MeV is used in the figure. Similar to
Fig. 2, one does not find the $m_{\perp}$-scaling law in these radius
parameters.

\end{description} 

\end{document}